\documentclass[doublecol]{epl2} 

\title{Electron pockets and pseudogap asymmetry observed in the thermopower of underdoped cuprates.}
\shorttitle{Electron pockets and pseudogap asymmetry in the thermopower of cuprates.} 

\author{J.G. Storey\inst{1,2} \and J.L. Tallon\inst{2} \and G.V.M. Williams\inst{1}}
\shortauthor{J.G. Storey \etal}

\institute{                    
  \inst{1} School of Chemical and Physical Sciences, Victoria University - 
P.O. Box 600, Wellington, New Zealand\\
  \inst{2} MacDiarmid Institute - Callaghan Innovation,
P.O. Box 31310, Lower Hutt, New Zealand
}

\abstract{
We calculate the diffusion thermoelectric power of high-$T_c$ cuprates using the resonating-valence-bond spin-liquid model developed by Yang, Rice and Zhang (YRZ). In this model, reconstruction of the energy-momentum dispersion results in a pseudogap in the density of states that is heavily asymmetric about the Fermi level. The subsequent asymmetry in the spectral conductivity is found to account for the large magnitude and temperature dependence of the thermopower observed in underdoped cuprates. In addition we find evidence in experimental data for electron pockets in the Fermi surface, arising from a YRZ-like reconstruction, near the onset of the pseudogap in the slightly overdoped regime.
}

\begin{document}

\maketitle

\section{Introduction}
A central issue in the quest to understand the high-$T_c$ cuprate superconductors is the nature and origin of the pseudogap, a depletion in the density of states, that dominates thermodynamic and transport properties across half of the temperature-doping phase diagram. The two heavily-debated opposing viewpoints are that it is either due to phase incoherent electron pairing, or instead arises from some competing order. The presence or absence of particle-hole symmetry is of vital importance in this debate. Superconducting excitations exhibit symmetry about the Fermi surface, whereas competing orders can have their locus of excitations along other parts of the Brillouin zone such as the antiferromagnetic zone boundary.  Recent angle-resolved photoemission spectroscopy (ARPES) studies have detected particle-hole asymmetric spectra in the pseudogap phase\cite{YANG1,HASHIMOTO1}. This was missed by earlier ARPES studies due to the widespread practice of symmetrizing the spectra to aid analysis. 

A related topic that has become of considerable interest recent years is the possibility of electron pockets in the Fermi surface of underdoped high-$T_c$ cuprates. This has been spawned by quantum oscillations\cite{DOIRON,SEBASTIAN}, Hall effect\cite{LEBOEUF} and thermopower\cite{LALIBERTE} measurements of strongly underdoped samples deep in the pseudogap regime near $x=1/8^{th}$ doping, where stripe order is inferred to be present. Direct proof of the existence and location of these pockets in momentum space would put significant constraints on the origin of the mysterious pseudogap and possibly the origin of superconductivity in these materials. In contrast, several Fermi surface reconstruction models predict electron pockets appearing with the onset of the pseudogap in the slightly overdoped regime ($x\approx0.19$) before disappearing at lower dopings. 

One such model is the resonating-valence-bond spin-liquid model developed by Yang, Rice and Zhang (YRZ)\cite{YRZ}.
It has achieved considerable success in describing several measurable properties of the high-$T_c$ cuprate superconductors. Some examples include the specific heat\cite{LEBLANC1,BORNE2}, Raman spectra\cite{VALENZUELA,LEBLANC2}, penetration depth\cite{CARBOTTE} and angle-resolved photoemission spectra\cite{YANGK2,YANG2}. A principal feature of this model is a self-energy term $E_g^2(\textbf{k})/(\omega+\xi_\textbf{k}^0)$ that reconstructs the energy-momentum dispersion into two branches, leading to a particle-hole asymmetric pseudogap in the density of states\cite{BORNE1} (see fig.~\ref{FIG1}(b)). At the same time, the large hole-like Fermi surface is transformed into small nodal hole pockets and, for small values of $E_g$, antinodal electron pockets (see fig.~\ref{FIG1}(a)). Recently we showed that this reconstruction reproduces the two-component electronic behaviour observed in NMR Knight shift experiments\cite{STOREYNMR}.
As a further test of the YRZ ansatz we calculate the diffusion thermoelectric power (TEP), which provides a direct measure of particle-hole asymmetry in the spectral conductivity. From these calculations we identify features in the TEP corresponding to electron pockets in the Fermi surface appearing near the onset of the pseudogap in the slightly overdoped regime.

\section{Formalism}
\label{sec:Formalism}

Detailed descriptions of the YRZ model have been published several times\cite{YRZ,SCHACHINGER,BORNE1}, but for completeness we briefly list the equations used in this work.
In the normal state the coherent part of the electron Green's function is given by
\begin{equation}
G(\textbf{k},\omega,x)=\frac{g_t(x)}{\omega-\xi_\textbf{k}-\frac{E_g^2(\textbf{k})}{\omega+\xi_\textbf{k}^0}}
\label{eq:GYRZ}
\end{equation}
where $\xi_\textbf{k}=-2t(x)(\cos k_x+\cos k_y)
-4t^\prime(x)\cos k_x\cos k_y
-2t^{\prime\prime}(x)(\cos 2k_x+\cos 2k_y)-\mu_p(x)$ is the tight-binding energy-momentum dispersion, $\xi_\textbf{k}^0=-2t(x)(\cos k_x+\cos k_y )$ is the nearest-neighbour term, and $E_g(\textbf{k})=[E_g^0(x)/2](\cos k_x-\cos k_y)$
is the pseudogap with $E_g^0(x)=3t_0(0.2-x)$ for $x\leq 0.2$, while for $x>0.2$ $E_g^0(x)=0$. Here we take the closure of the pseudogap to lie at $x=0.2$ in continuity with YRZ, however we have extensively shown this to occur at slightly lower doping $x=0.19$\cite{OURWORK1}. The chemical potential $\mu_p(x)$ is chosen according to the Luttinger sum rule, with the values used in this work listed in Table~\ref{MUTABLE}.
The doping-dependent coefficients are given by $t(x)=g_t(x)t_0+(3/8)g_s(x)J\chi$, $t^\prime(x)=g_t(x)t_0^\prime$ and $t^{\prime\prime}(x)=g_t(x)t_0^{\prime\prime}$, where $g_t(x)=2x/(1+x)$ and $g_s(x)=4/(1+x)^2$ are the Gutzwiller factors. The bare parameters $t^\prime/t_0=-0.3$, $t^{\prime\prime}/t_0=0.2$, $J/t_0=1/3$ and $\chi=0.338$ are the same as used previously\cite{YRZ}.

Equation~\ref{eq:GYRZ} can be re-written as
\begin{equation}
G(\textbf{k},\omega,x)=\sum_{\alpha=\pm}{\frac{g_t(x)W_\textbf{k}^\alpha(x)}{\omega-E_\textbf{k}^\alpha(x)}}
\label{eq:GYRZ2}
\end{equation}
where the energy-momentum dispersion is reconstructed by the pseudogap into upper and lower branches
\begin{equation}
E_\textbf{k}^\pm=\frac{1}{2}(\xi_\textbf{k}-\xi_\textbf{k}^0)\pm\sqrt{\left(\frac{\xi_\textbf{k}+\xi_\textbf{k}^0}{2}\right)^2+E_g^2(\textbf{k})}
\label{eq:EK}
\end{equation}
that are weighted by
\begin{equation}
W_\textbf{k}^\pm=\frac{1}{2}\left[1\pm\frac{(\xi_\textbf{k}+\xi_\textbf{k}^0)/2}{\sqrt{[(\xi_\textbf{k}+\xi_\textbf{k}^0)/2]^2+E_g^2(\textbf{k})}}\right]
\label{eq:WK}
\end{equation}
The spectral function is given by
\begin{equation}
A(\textbf{k},\omega,x)=\sum_{\alpha=\pm}{g_t(x)W_\textbf{k}^\alpha\delta(\omega-E_\textbf{k}^\alpha)}
\label{eq:Akw}
\end{equation}
from which the density of states (DOS) can be calculated
\begin{equation}
N(\omega)=\sum_{\textbf{k}}{A(\textbf{k},\omega)}
\label{eq:DOS}
\end{equation}
The Fermi surface given by $E^{\pm}_\textbf{k}=0$, and DOS are plotted for several values of $x$ in figs.~\ref{FIG1}(a) and (b), respectively.

\begin{table}
\caption{Values of $\mu_p(x)$ used in this work.}
\label{MUTABLE}
\vspace{10pt}
\centerline{
\begin{tabular}{|c|c||c|c|}
\hline
$x$ & $\mu_p/t_0$ & $x$ & $\mu_p/t_0$\\
\hline 
0.10 & -0.250 & 0.18 & -0.380\\
0.12 & -0.270 & 0.19 & -0.420\\
0.14 & -0.295 & 0.20 & -0.460\\
0.16 & -0.315 & 0.25 & -0.588\\
\hline
\end{tabular}}
\end{table}

\section{Thermopower}
\begin{figure}
\centering
\includegraphics[width=70mm,clip=true,trim=0 0 0 0]{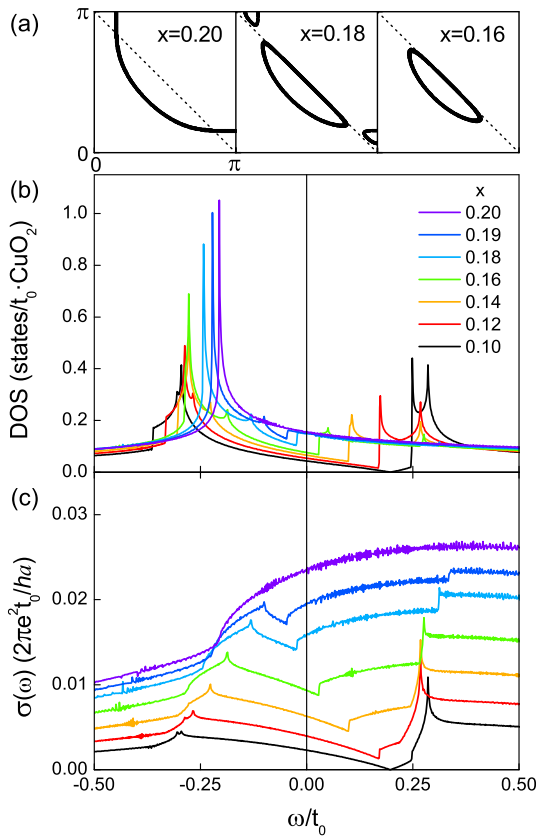}
\caption{(Color online) (a) Evolution of the Fermi surface with decreasing doping in the first quadrant of the Brillouin zone as described by the YRZ model. (b) Density of states and (c) conductivity calculated from the YRZ model, where $a$ is the lattice spacing. The pseudogap arising from the reconstruction of the energy-momentum dispersion is strongly asymmetric about the Fermi level $\omega=0$. For $x$ = 0.18 and 0.19 the crossing of the upper branch of the dispersion results in electron pockets near ($\pi$,0) and (0,$\pi$), and a pseudogap below $E_F$.}
\label{FIG1}
\end{figure}

\begin{figure}
\centering
\includegraphics[width=70mm,clip=true,trim=0 0 0 0]{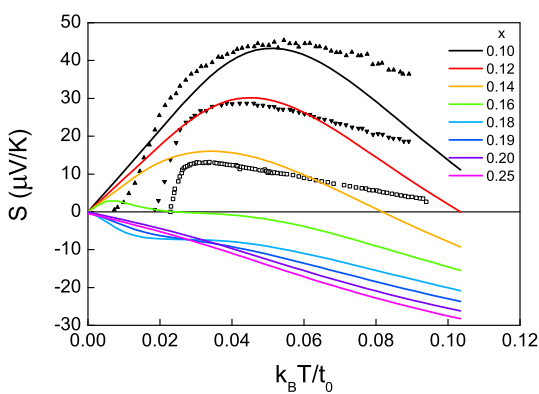}
\caption{(Color online) Thermopower calculated from the YRZ spectral function eq.~\ref{eq:Akw}. The large positive values, comparable with experiment, originate from the pseudogap-induced asymmetry in the conductivity. The undulation in the $x$ = 0.18 and 0.19 curves is a signature of electron pockets in the Fermi surface. Experimental Bi-2212 data, with $T$ rescaled by $t_0=0.3eV$, are from Munakata \textit{et al}.\cite{MUNAKATA} (filled triangles), and Fujii \textit{et al}.\cite{FUJII} (open squares).}
\label{FIG2}
\end{figure}

The diffusion thermopower is given by\cite{ALLEN}
\begin{equation}
S(T)=\frac{1}{\left|e\right|T\sigma_{dc}(T)}\int_{-\infty}^{\infty}{\sigma\left(\omega\right)\omega\left(\frac{\partial{f}}{\partial{\omega}}\right)d\omega}
\label{TEPEQ}
\end{equation}
where $f$ is the Fermi function. $\sigma_{dc}$ is the dc conductivity given by
\begin{equation}
\sigma_{dc}(T)=\int_{-\infty}^\infty{\sigma(\omega)\left(-\frac{\partial f}{\partial \omega}\right)d\omega}
\label{eq:SIGMAT}
\end{equation}
and $\sigma(\omega)$ is the spectral conductivity. The TEP as defined by eq.~\ref{TEPEQ} is a measure of the asymmetry in $\sigma(\omega)$ about $\omega=0$ via the spectral window $\omega(\partial{f}/\partial{\omega})$. Approximating vertex corrections by $v_x=\partial{\xi_\textbf{k}}/\partial{k_x}$,  $\sigma(\omega)$ is calculated as\cite{PRUSCHKE,HILDEBRAND}
\begin{equation}
\sigma(\omega)=\frac{e^2}{V}\frac{2\pi}{N}\sum_\textbf{k}{v_x^2A^2(\textbf{k},\omega)}
\label{eq:SIGMA2}
\end{equation}
A more complete treatment of vertex corrections can be found in refs.~\cite{KONTANI,HARTNOLL}. Note that eq.~\ref{eq:SIGMA2} is not the same as the optical conductivity, which is a convolution of the spectral conductivity above and below $\omega$=0\cite{ILLES,DENOBREGA}. Substituting the YRZ spectral function given by eq.~\ref{eq:Akw} produces the conductivity and TEP curves shown in fig.~\ref{FIG1}(c) and fig.~\ref{FIG2}.

The TEP of the high-temperature cuprate superconductors was early shown\cite{OBERTELLI} to exhibit universal behaviour. Roughly speaking, the linear part of the TEP takes the form $S(T)=S_0-\alpha T$. $S_0$ is positive and large in the underdoped regime and decreases with doping, becoming negative in the overdoped regime, while $\alpha$ is approximately doping independent. The TEP calculated from the YRZ spectral function reproduces the large magnitude and $T$-dependence observed in underdoped cuprates. This is due to the strong asymmetry in $\sigma$ about $\omega$=0 arising from the pseudogap. A similar conclusion was previously reached by Hildebrand \textit{et al}.\cite{HILDEBRAND} from calculations based on the fluctuation-exchange approximation applied to the Hubbard model. When the pseudogap is omitted, as in Kondo \textit{et al.}\cite{KONDOTEP}, the calculated diffusion TEP fails to reproduce the large values observed in the underdoped region.

An undulation in the TEP appears for $x$ = 0.18 and 0.19. At these dopings, both branches of the reconstructed dispersion cross the Fermi level ($E_F$), resulting in electron pockets near the ($\pi$,0) points on the Fermi surface (fig.~\ref{FIG1}(a)), and a pseudogap that lies entirely below $E_F$ (fig.~\ref{FIG1}(b)). A residual undulation is also present for $x$ = 0.16 due to the close proximity of the upper branch to $E_F$. We address experimental observations of this undulation later - a key result of this paper. For $x$=0.2 and above, the TEP is negative with a linear $T$-dependence down to $T$=0. But unlike experimental data $S_0$ does not become negative, and the slope $\alpha$ increases. This originates from the broad section of positive slope in $\sigma(\omega)$ through $\omega$=0.

A better match with data in the overdoped regime is found using the Boltzmann expression for $\sigma(\omega)$ given by
\begin{equation}
\sigma(\omega)=\frac{e^2}{V}\sum_{\textbf{k}}{v_x(\omega,\textbf{k})\ell_x(\omega,\textbf{k},T)\delta[\xi_\textbf{k}-\omega]}
\label{SIGMAEQ}
\end{equation}
$\ell_x$ is the mean free path given by
\begin{equation}
\ell_x=v_x(\omega,\textbf{k})\cdot\tau(\omega,\textbf{k},T)
\label{MFPEQ}
\end{equation}
and $\tau(\omega,\textbf{k},T)$ is the relaxation time.
Assuming a constant mean free path, Kondo \textit{et al}. calculated the TEP from the measured ARPES dispersions of a series of Bi-2201 samples\cite{KONDOTEP}. 
Good correspondence with measured TEP data was achieved for over- and optimally doped samples without any free parameters.
In this case, $\sigma(\omega)\propto\sum_{\textbf{k}}{v_x\delta[\xi_\textbf{k}-\omega]}$ is effectively a velocity-weighted DOS. It is asymmetric about $E_F$, peaking at $E_{vHs}$ due to the saddle-point van Hove singularity (vHs) in the DOS. With increasing doping, $E_F$ approaches and crosses the vHs, resulting in a vanishing positive peak in the TEP, and a change in $S_0$ from positive to negative values. 

The constant-mean-free-path assumption implies that the scattering rate $\tau^{-1}_\textbf{k}\propto v_\textbf{k}$. Since $v_\textbf{k}$ is smallest at the saddle-points of the dispersion located at ($\pm\pi$,0) and (0,$\pm\pi$), $\tau^{-1}(\omega)$ goes roughly as $|\omega-E_{vHs}|$. Supporting evidence for a small scattering rate at the saddle points comes from ARPES measurements of the lifetime of Bloch states at $E_F$\cite{KONDOTAU}, as well as the detection of sharp quasiparticle peaks at ($\pi$,0) in overdoped Bi-2201\cite{YANGK}, Tl-2201\cite{PEETS} and Bi-2212\cite{YUSOF} where $E_F$ is close to the vHs. 

\begin{figure}
\centering
\includegraphics[width=70mm,clip=true,trim=0 0 0 0]{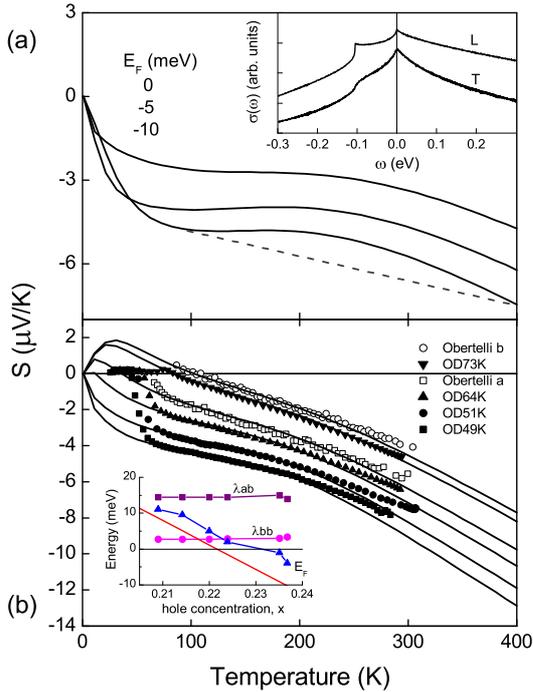}
\caption{(Color online) (a) The thermopower calculated from the bilayer $\epsilon_k$-dispersion of Bi-2212 assuming a constant mean free path. The Fermi level, $E_F$, is measured relative to the antibonding band van-Hove singularity. The curvature above 100K is due to the bonding band vHs. Inset: $\sigma(\omega)$ calculated with a constant mean free path (L), and the model scattering rate given by eq.~\ref{TAUEQ} (T). (b) Fits to our own Bi$_{1.8}$Pb$_{0.3}$Sr$_{1.9}$CaCu$_2$O$_8$ TEP data (black solid symbols) and data from Obertelli \textit{et al}.\cite{OBERTELLI}, calculated from the bilayer dispersion with the model scattering rate. Inset: Doping dependence of the $E_F$ and coupling constants extracted from the fits. For comparison the red line shows the $E_F$ determined from the electronic entropy\cite{STOREYENTROPY}.
}
\label{BILAYERFIG}
\end{figure}

To further check the validity of this assumption we have extended the constant-mean-free-path approach to calculate the TEP of Bi-2212, using a bilayer ARPES dispersion\cite{STOREYPHASE} that we have previously shown accurately describes the electronic entropy\cite{STOREYENTROPY,STOREYPG} and superfluid density\cite{STOREYENTROPY}.
The calculated TEP is shown in fig.~\ref{BILAYERFIG}(a) with comparative overdoped
Bi$_{1.8}$Pb$_{0.3}$Sr$_{1.9}$CaCu$_2$O$_8$ data shown in panel (b). The overall doping- and $T$-dependence is successfully reproduced, confirming the validity of the constant-mean-free-path assumption in the overdoped regime. The calculations reveal that the curvature above 100K originates from the bonding band vHs, located approximately 100meV below the antibonding band vHs, which produces an additional peak in $\sigma(\omega)$ evident in the curve marked ``L'' in the inset to fig.~\ref{BILAYERFIG}(a). The same curvature is visible in the overdoped data of Munakata \textit{et al}\cite{MUNAKATA}.

\begin{figure}
\centering
\includegraphics[width=70mm,clip=true,trim=0 0 0 0]{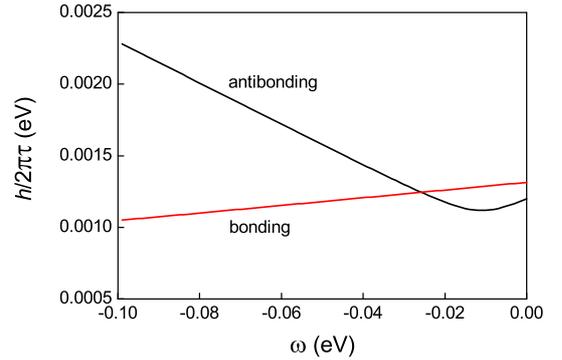}
\caption{(Color online) Bonding and antibonding band scattering rates calculated from eq.~\ref{TAUEQ} for $p$=0.209 and $T$=30K.
}
\label{TAUFIG}
\end{figure}

In order to obtain a more quantitative comparison with the data we adopt a scattering rate of the form
\begin{equation}
\hbar\tau^{-1}\left(\omega,T\right)=\lambda\sqrt{\left(\pi{}k_BT\right)^2+\left(\omega-E_{vHs}\right)^2}+a
\label{TAUEQ}
\end{equation}
where $\lambda$ is a coupling constant. The parameter $a$ takes a fixed value of 1meV and is included to prevent $\tau$ from diverging at the vHs.
The square root term in eq.~\ref{TAUEQ} is similar to the implementation of the max($|\omega|,T$) marginal Fermi liquid single-particle scattering rate employed by Abrahams and Varma\cite{ABRAHAMS}. $\lambda$ provides an adjustable parameter with which the sharpness of the peaks in $\sigma(\omega)$ can be tuned. Separate instances of eq.~\ref{TAUEQ} were applied to the antibonding and bonding bands with coupling constants $\lambda_{AB}$ and $\lambda_{BB}$  respectively. The inset to fig.~\ref{BILAYERFIG}(b) shows the doping dependence of these parameters, and the location of $E_F$ relative to the antibonding band vHs extracted from the fits.
$\lambda_{AB}$ and $\lambda_{BB}$ are approximately independent of doping over the range studied. The scattering rates for the $x$=0.209 sample are shown in fig.~\ref{TAUFIG}. At the Fermi level, the bonding band scattering rate is larger than that of the antibonding band, while at higher binding energies the scattering rates cross each other and the situation reverses. These features have been observed in the ARPES-derived scattering rates of Bi-2212\cite{BORISENKO2} and Y-123\cite{BORISENKO}, and have been attributed to collective spin excitations.
The doping dependence of $E_F$ is consistent with the Fermi level crossing the antibonding vHs
between $p$=0.22 and 0.23 ($T_c\approx$ 60K) as already deduced from ARPES\cite{2212VHS} and our previous fits to
the electronic entropy\cite{STOREYENTROPY}. 

\begin{figure}
\centering
\includegraphics[width=70mm,clip=true,trim=0 0 0 0]{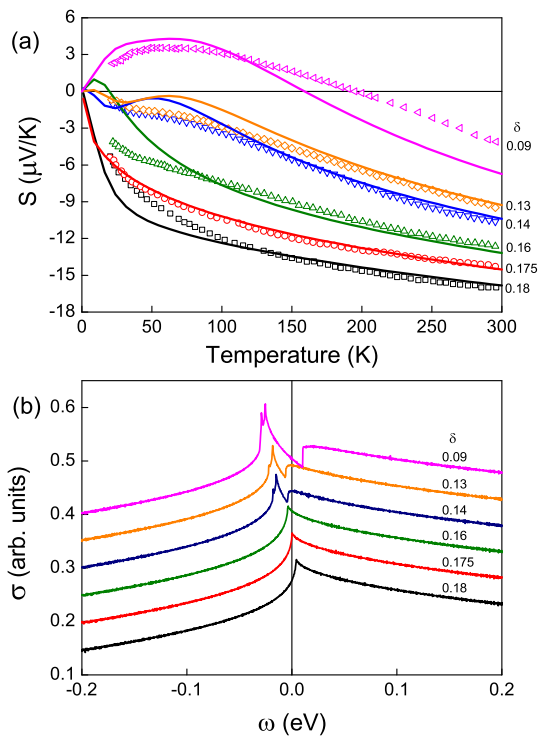}
\caption{(Color online) (a) TEP of Bi$_2$Sr$_2$CuO$_{6+\delta}$\cite{KONSTANTINOVIC2} and fits calculated from the energy-momentum dispersion incorporating Fermi surface reconstruction and assuming a constant mean free path. The undulation in the fits to the $\delta$=0.13 and 0.14 curves originates from electron pockets in the Fermi surface. (b) $\sigma(\omega)$ corresponding to the calculated curves in (a) showing the evolution of the pseudogap.
}
\label{KONSTFIG}
\end{figure}

In Bi-2212 the pseudogap opens as doping is reduced below $x_{crit}$=0.19\cite{ENTROPYDATA2,STOREYENTROPY}. But an undulation in the TEP associated with electron pockets is absent in both the literature and in our own underdoped data, probably because it is masked by the high value of $T_c$ and the onset of superconducting fluctuations above $T_c$.
However, data showing an undulation has been reported several times in the TEP of single-layer Bi-2201\cite{SMITS,SUBRAMANIAM,KONSTANTINOVIC2,OKADA,OKADA2} and in Tl-1201\cite{SUBRAMANIAM} where $T_c\leq50$K. 
Figure \ref{KONSTFIG}(a) shows the Bi$_2$Sr$_2$CuO$_{6+\delta}$ data of Konstantinovi\'c \textit{et al.}\cite{KONSTANTINOVIC2} in which an undulation is observed for $\delta$=0.13 and 0.14. Overlaid on this data are curves calculated by applying a YRZ-like reconstruction to the energy-momentum dispersion of Bi-2201 from ref.~\cite{KONDOTEP}, and assuming a constant mean free path. We note that these fits are not specific to the YRZ model and we have found that, due to the small size of $E_g$ in this regime, identical results can be obtained from the standard antiferromagnetic Brillouin-zone-folding reconstruction model. The corresponding $\sigma(\omega)$ curves are shown in fig.~\ref{KONSTFIG}(b) illustrating firstly the traversal of the vHs through $E_F$, the subsequent opening of the pseudogap below $E_F$, and the eventual spanning of $E_F$ by the pseudogap as its magnitude increases with reducing doping. In principle, more precise fits could be made by introducing a more complex scattering rate, however the overall qualitative picture would remain unchanged.

\begin{figure}
\centering
\includegraphics[width=68mm,clip=true,trim=0 0 0 0]{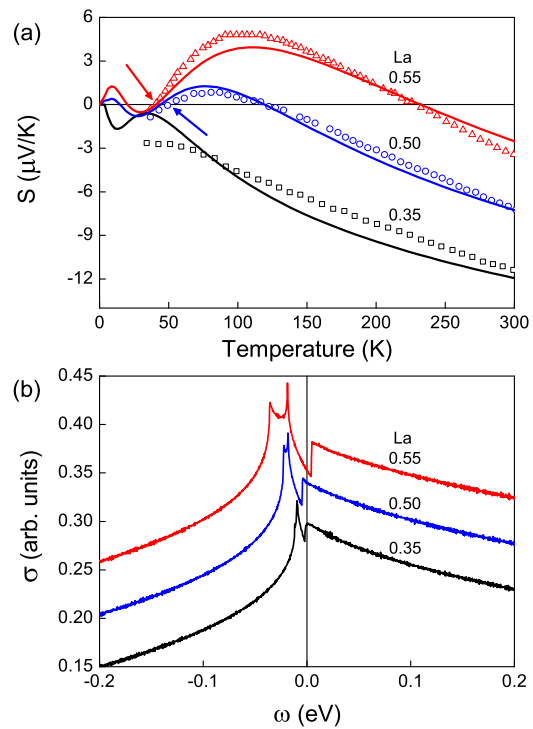}
\caption{(Color online) (a) TEP of Bi$_2$Sr$_{2-x}$La$_x$CuO$_y$\cite{OKADA} and fits calculated from the energy-momentum dispersion incorporating Fermi surface reconstruction and assuming a constant mean free path. Electron pockets at or near $E_F$ can be inferred from the TEP going to zero above $T_c$ (arrowed). (b) $\sigma(\omega)$ corresponding to the calculated curves in (a) showing the evolution of the pseudogap.
}
\label{OKADAFIG}
\end{figure}

\begin{figure}
\centering
\includegraphics[width=70mm,clip=true,trim=0 0 0 0]{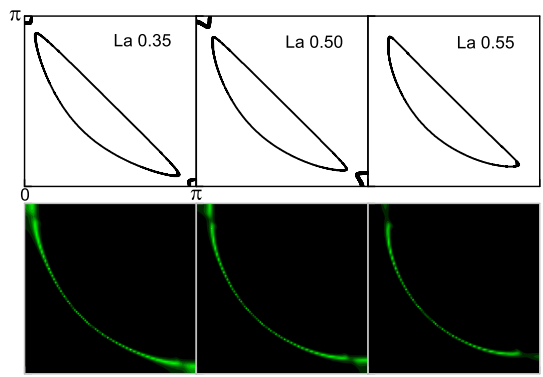}
\caption{(Color online) The Fermi surface and spectral functions corresponding to the calculated TEP curves in fig.~\ref{OKADAFIG}. The spectral functions are calculated using a Lorentzian broadening with a full-width at half-maximum of 10meV.
}
\label{OKADAFSFIG}
\end{figure}

In fig.~\ref{OKADAFIG} we show similar fits to La-doped Bi-2201 data from Okada \textit{et al}\cite{OKADA}, with corresponding Fermi surfaces and spectral functions shown in fig.~\ref{OKADAFSFIG}. Another signature of electron pockets at, or close to $E_F$, is the decrease in TEP to zero above $T_c$, near 50K (arrowed). It should be noted that $T_c^{max}$ is approximately 35K in this system\cite{OKADA2}. By contrast, in the absence of electron pockets or Fermi surface reconstruction, the calculated TEP decreases approximately linearly to zero at $T=0$ as shown in figs.~\ref{BILAYERFIG}(b) \& \ref{KONSTFIG}(a) and ref.~\cite{KONDOTEP}. These signatures of electron pockets may be rendered more evident at low temperature using high magnetic fields or zinc substitution.

\section{Summary}
In summary, we have calculated the diffusion thermopower of underdoped cuprates from the resonating-valence-bond spin-liquid model developed by Yang, Rice and Zhang. The pseudogap arising from reconstruction of the Fermi surface in this model is asymmetric about the Fermi level, and can account for the large positive magnitude and temperature dependence of the TEP observed experimentally. In the overdoped regime, the evolution of the $T$-dependence with increasing doping is accurately described in terms of the changes in electronic structure observed by photoemission, in particular the Fermi level crossing the vHs, combined with a constant mean free path. For Bi-2212, the calculations reveal effects of bilayer splitting in the data. Finally, we have identified an undulation in the $T$-dependence at low temperatures that can be attributed to the presence of electron pockets in the Fermi surface near the onset of the pseudogap in the slightly overdoped regime. This feature is consistent with both YRZ and antiferromagnetic Brillouin-zone-folding Fermi surface reconstruction models. These findings suggest that the diffusion component dominates the TEP over any phonon drag contribution. This contrasts with the approach taken earlier by Trodahl\cite{PHONONDRAGMODEL}, who modelled the TEP as the sum of a negative linear metallic diffusion component, and a positive phonon drag component that rises at low temperatures and saturates at high temperatures. Such a model does not include the intricacies of the electronic structure that have subsequently been found to exist in these materials - details which the TEP is able to evidently expose.

\bibliographystyle{eplbib}

\end{document}